# Improved Flow Recovery from Packet Data

Challenges in Accurate Flow Recovery from Packet Datasets

Anthony Kenyon,
Hyperscalar Ltd, Farnham, UK
tony.kenyon@ieee.org

David Elizondo
Institute of Artificial Intelligence
School of Computer Science and Informatics,
De Montfort University, UK.
elizondo@dmu.ac.uk

Lipika Deka*
Institute of Artificial Intelligence
School of Computer Science and Informatics,
De Montfort University, UK.
lipika.deka@dmu.ac.uk
*corresponding author

*Abstract*— Typical event datasets such as those used in network intrusion detection comprise hundreds of thousands, sometimes millions, of discrete packet events. These datasets tend to be high dimensional, stateful, and time-series in nature, holding complex local and temporal feature associations. Packet data can be abstracted into lower dimensional summary data, such as *packet flow records*, where some of the temporal complexities of packet data can be mitigated, and smaller well-engineered feature subsets can be created. This data can be invaluable as training data for machine learning and cyber threat detection techniques. Data can be collected in real-time, or from historical packet trace archives. In this paper we focus on how flow records and summary metadata can be extracted from packet data with high accuracy and robustness. We identify limitations in current methods, how they may impact datasets, and how these flaws may impact learning models. Finally, we propose methods to improve the state of the art and introduce proof of concept tools to support this work.

I. INTRODUCTION

Access to good datasets with which to build computational intelligence models, particularly labelled data, is a major challenge in machine learning today. In fields such as cybersecurity, network analysis and traffic engineering, packet data is an invaluable resource for raw low-level network events. This data can provide critical information on bandwidth utilisation, network congestion, timing and latency, protocol distributions, as well as threat and anomaly patterns. Network data is collected using a variety of methods [ZHOU18] and be presented in a variety of formats [KEN20]. Some of the most popular formats, include:

- PCAP – used for lossless binary full packet trace capture of timestamped events.
- CSV - commonly used to summarise PCAP traces as packet flow records or metadata summaries.
- XML and JSON – used in a similar way to CSV, but as structured objects, events and metadata

Most flow data is recorded in real time, using networked systems that support flow standards such as NetFlow [KERR01] and IPFIX [RFC7011][1], which we discuss later in this paper[2].

---

[1] Derived from Netflow version 9
[2] Additional flow standards are largely based on the original Cisco NetFlow standard, and include jFlow (developed by Juniper networks), NetStream

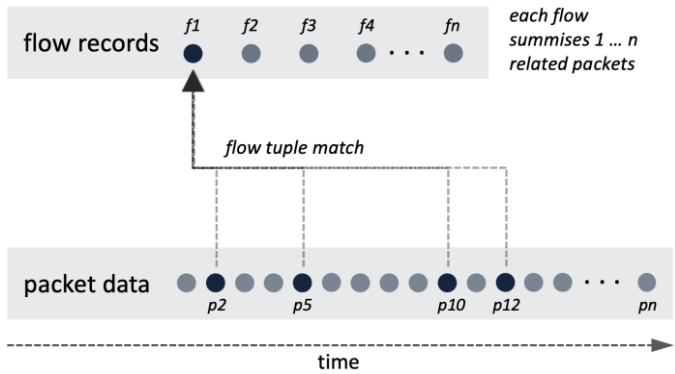

Fig. 1. Illustrates the many-to-one relationship between packet and flow data. Packets are sequentially ordered and timestamped. Each packet may belong to a discrete flow record, but not more than one flow record. Each flow record also has a timestamp, typically mapped to the first packet's timestamp recorded in the flow. Note that flow order depends on the flow recovery context; order should strictly follow timestamp order. In practice flows may appear out of order in the flow dataset (for example if flows are being recovered *ex post facto* then they may be exported in the order that flows are terminated in memory).

Packet datasets tend to be large[3], often spanning throughout multiple days. Due to their size and inherent complexity most modern PCAP datasets tend to be accompanied by flow summaries. Flow records are a widely used method to compress packet data, reducing much of the dimensional and temporal complexity, and reducing the features space down to useful subsets through careful feature engineering - see Fig.1.

The ability to retrospectively recover flow data from existing packet datasets (particularly labelled data) can be invaluable in fields such as cybersecurity, since this data can be remodelled as training data for machine learning and predictive analytics. Furthermore, as new cyber threats emerge it may be useful to engineer new features from original PCAP data, to test out new intrusion detection techniques. This is a complex technical challenge and there are few tools available that can extract flow summaries retrospectively with high fidelity. Such tools that do exist may also exhibit design flaws that can compromise data utility [KEN20].

Reconstructing flow summaries from packet data is not an exact science, and there are multiple challenges, inconsistencies and ambiguities that need to be tackled, including the need to track protocol interaction statefully. This paper examines the state of the art in flow recovery techniques, highlighting the key challenges in the recovery process, and offering a methodology for improving flow record accuracy.

## II. BACKGROUND

As introduced above, perhaps the most useful way we can reduce dimensionality in packet data is by mapping packets that are associated over time into higher-level metadata containers called *flows*.

### A. What is a flow

The term *packet flow* is used quite loosely in the literature. In practice there is no uniformly accepted definition[4], however it is essentially characterised as a sequence of related packets, identified by a common 5-tuple identifier, including:

- Source and Destination IP address
- Source and Destination Port number
- Protocol type (e.g., TCP)

A flow represents a structured service interaction between systems, using an established set of common resources, for a limited time - for example sending a file between two systems, or posting an email.

### B. Flow standards

In the more formal sense, there are industry and international standards dating back to the 1990s [RFC1272, RFC2721] for flow encoding and representation, resulting in the subsequent development of NetFlow [KERR01] and IPFIX [RFC7011]. In practice this means that networked systems are capable of capturing flow features in a standard, extensible manner. Detailed metadata on flow statistics, state and timing can be collated, as well as specific flow behavioural aspects (using techniques such as *Deep Flow Inspection* (DPI) [LIU14, ZHOU18]).

---

(developed by Huawei) and cflowd (developed by Alcatel-Lucent). A related standard, sFlow, is based uses sampling techniques.
[3] Potentially hundreds of gigabytes in size [KEN20].

[4] This concept is normally associated with the TCP/IP protocol suite, although this can be abstracted to other protocol types with a similar architecture.

*C. Recovering flows from live and archived data*

While flows can be captured in real time [ZHOU18], the ability to reconstruct flows from live or archived PCAP traces, is a valuable way to mine raw event data, particularly for use in machine learning and security research (for example intrusion detection or DDoS research).

For older published datasets. this also allows us to look back in time, using consistent modern flow extraction techniques, engineering new features that allow us to compare new learning techniques with established models in the literature. Such data is typically hard to collect [KEN20], and the data may be highly specialised to the deployment context. In such cases these raw packet archives may contain invaluable or rare event data. Flow reconstruction allows us to correct for any design flaw or errors in original summary data, as well as introduce new feature subsets for use in learning models. By recovering flow data, we can also re-evaluate the quality of such datasets against modern standards, as well as identifying changes in protocol distributions, traffic volumes and threat patterns over time. For example, we can see a clear shift towards encrypted web traffic in the changes in distribution of HTTP and HTTPS over time, and some services disappear altogether).

*D. Benefits of using flows*

The key benefits of using flows are significant, since flows aggregate related packet events across time, as well as compressing overall feature dimensionality. In addition:

- Memory reduction is significant. It is possible to process much larger sample spaces over longer time periods due to this compression.

- Whilst we lose some fidelity in event timing, we can engineer features that statistically characterise flows in highly useful ways (start and end time, packet intervals and variances, by flow direction etc.). We can also model burst characteristics, variance in timing and latency, jitter etc. These summary features may have more predictive power and be better suited for use in machine learning, analytics and anomaly detection, than raw packet events.

- Rich metadata can be captured on many other quantitative and state measures, such as flag counts, bytes and packets sent and received, statistics on maximum, minimum, average, variance and standard deviation for bytes and packet lengths sent and received.

- Payload information can be reconstructed across a flow and summarised statistically in a highly compact form (e.g., size by direction, content type, entropy etc.), without compromising data privacy.

- We can filter out highly correlated features or features with low variance, if desired.

By summarising packet data as flow records, we can provide high quality training data for machine learning and network analytics, without the burden of orchestrating, storing and processing millions of discrete events.

*E. Existing tools and data quality*

Today there are few open-source tools available that provide flow recovery, with detailed feature engineering suited to large-scale training dataset creation[5]. In the literature, various tools and scripts have been distributed with flow datasets, designed to produce flow or metadata summaries from packet data. Most of these tools are limited in functionality and are not archived or supported consistently [KEN20]. Arguably the most sophisticated tool for flow data engineering is a Java based package called CICFlowMeter [LASH13, HAB17], released by the University of New Brunswick (UNB) [SHAR18]. This is an excellent tool used to create many of the UNB flow summaries associated with their public datasets [KEN20]. Details on flow metadata output by Version 4 of the CICFlowMeter tool are described in [LASH18].

In examining packet traces and flows produced by existing tools, several artefacts reveal inconsistencies and subtle issues in flow recovery methods[6], which manifest themselves as:

---

[5] A number of tools exist, including nProbe and ntopng, YAF, SiLK, Joy, FlwBAT, NetFlow Analyzer, Bro, and Suricata, but most are not ideally suited to large dataset generation.

[6] We identified multiple inconsistencies using statistical analysis of flows with, for example, anomalous FIN flag counts, and anomalous TCP and UDP source port allocations (where port identities appear that are normally reserved for destination services). In some cases, we observed incorrect source port allocation rates in excess of 20% of cases.

- Incorrect flow counts (more or less flows depending on whether flow state was correctly interpreted. This might happen depending on the implementation, for example when dealing with unusual flow states during a DDoS attack.

- Missing flag counts (for example TCP FIN flag counts are zero, or do not balance with SYN flags as expected). This may occur if flow termination logic is flawed and cached flow data structures are aged out too soon (resulting in later packets being ignored, depending on the implementation).

- Incorrect port assignment. Well-known and registered TCP and UDP ports appear in the source port column (likely due to incorrect flow direction assignment on partial flows).

- Out of sequence flows. Since some flows last longer than others, flow start times may be out of sequence in the resulting flow trace, as these long-lasting flows were aged out of memory later in the flow extraction process.

It is also worth noting that very few tools provide holistic analysis on the flow composition and dataset structure, and there is little in the way of feature engineering on packet payload composition [KEN23E]. Given the potential for using such datasets in areas such as cyber threat research and machine learning, we re-examine the flow recovery process, and address some of these issues. First we begin by enumerating the challenges faced during flow recovery in the following section.

### III. FLOW RECOVERY CHALLENGES

Packet datasets comprise high-dimensional, complex time-series event data, and are frequently 'messy', with missing data, duplicate events, errors, and behavioural anomalies [KEN20]. When considering the task of flow recovery, we need to address a number of issues that complicate the recovery process, ranging from benign artefacts to intended malicious features. These features are summarised in Fig.2 and discussed in detail below.

| Type | Issue | FHL | TSQ |
|---|---|---|---|
| Issues with real work network data | Poor protocol implementation | ✓ | |
| | Dropped and duplicate packets | ✓ | ✓ |
| | Protocol or endpoint failures | ✓ | |
| | Flow State and Persistence | ✓ | |
| | Timestamp reliability | | ✓ |
| | Session termination ambiguity | ✓ | ✓ |
| | State machine complexity | ✓ | |
| Issues at the point of capture | Defining a consistent flow tuple | ✓ | |
| | Flow Direction and dynamics | ✓ | ✓ |
| | Partial Flow Capture | ✓ | ✓ |
| Adversarial or anomalous activity | Protocol abuse | ✓ | ✓ |
| | Protocol state machine abuse | ✓ | ✓ |
| | Persistent malicious flows | ✓ | ✓ |

Note: FHL = Flow Handling Logic. TSQ = Trainin Set Quality

Fig. 2. Summary of the main issues that need to be tackled in flow recovery and their potential impact on flow handling logic (FHL) and training set quality (TSQ).

#### A. Benign complications with real network data

Here we consider some key features of packet data that are to be anticipated in real network deployments:

- **Poor protocol implementation:** in the past there have been instances where protocol implementations have deviated from the agreed standard. This can happen for a number of reasons, such as misinterpretation of the standard, poorly defined or ambiguous specification in the standard, or through proprietary extensions and optimisations.

- **Dropped and duplicate packets**: most modern networks are 'best effort' -depending on load - and may not (at the physical layer) guarantee packet delivery. Packets might also be dopped by the network, interface card, or due to filtering, or may be missed by the collector - e.g., if alternate paths are chosen for certain traffic types. The captured data may therefore be missing key data, or may contain duplicates (e.g., TCP retransmissions, UDP data might be flooded down multiple paths).

- **Protocol or endpoint failures**: sometimes network systems fail, either at the hardware level or due to software bugs. This can produce unusual or anomalous packet behaviour within a packet trace.

- **Flow State and Persistence**: some flows may persist for long periods (hours, or even days).

Flow recovery must account for this when maintaining state.

- **Timestamp reliability**: Real networked systems of typically synchronised with a central time service. This becomes more complex in large enterprises, with a geographically distributed estate spanning multiple time zones. There may be synchronisation challenges here, as well as misconfigurations in time zone or time, in packet collectors and tools. When joining or correlating packet data this can be even more problematic. Changing packet timestamps post-facto is also non-trivial since this may require changes to checksums and embedded features, depending on the protocols used. Lastly, packet traces are normally in timestamp order, however flow records may not be in time order, for reasons we discussed earlier.

- **Session termination ambiguity**: in some protocols, such as TCP, session termination can be a somewhat ambiguous in real-world deployments, since it may depend on multiple exit criteria (e.g., with TCP this relies on an orchestrated FIN flag acknowledgement sequence, as well as timeouts, and it is possible that late or duplicate FIN flags may arrive post termination). Flow recovery must handle this statefully so that flags are accurately counted and that cached flow states are terminated unambiguously.

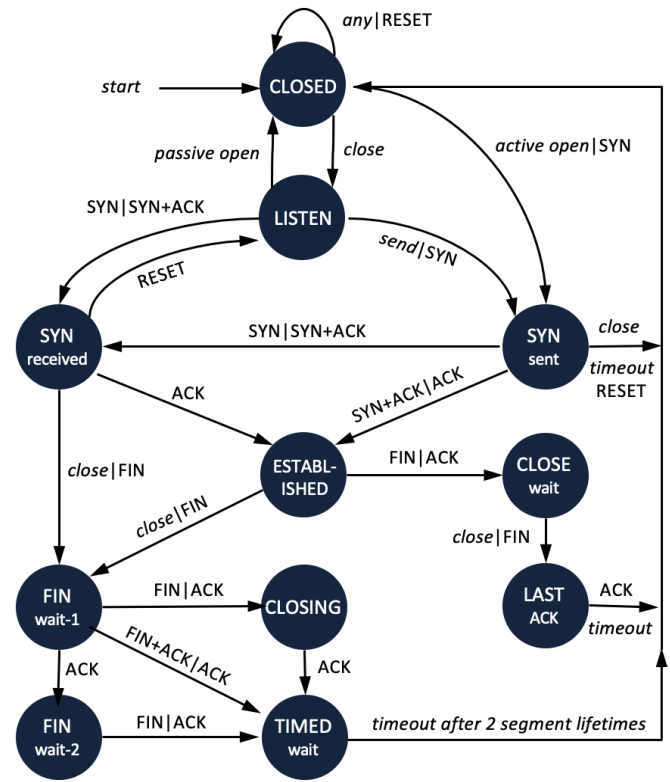

Fig. 3. TCP state machine, adapted from (Comer, 1995). State is clodely controlled through the use of special bit flags within the TCP header, exchanged in bi-directional packet flows. Where the data being sythesised is packet data, modelling state can be especially challenging with generative methods, and is likley to require adidtional helper methods.

- **State machine complexity**: protocols tend to be stateful or stateless, and this often depends on the position that a protocol occupies in the OSI model hierarchy [DAY83][7], and the deployment context. The majority of network traffic today is based on the Transmission Control Protocol/Internet Protocol (TCP/IP) protocol suite [WRIGHT95]. TCP/IP features two main transport layers, TCP and User Datagram Protocol (UDP) TCP is a *stateful* protocol, meaning that flow recovery must be capable of handling packets as a sequence of related events, with well-defined flow setup and teardown conditions, and *state transitions* - see Fig.3 and 5. When recovering such flows special processing logic is required to detect the start and end of a flow, as well as emulating timeout conditions. For stateless protocols, such as UDP we can assume the presence of a bidirectional flow by aggregating events based on identical flow tuples, and terminate these flows based on inactivity timers. Note that in this case this is a design choice, and a matter of convenience, since we either end up with a few large bidirectional UDP 'flows', or a separate flow per packet per direction - which can significantly distort and complicate the data.

---

[7] In practice we can have many combinations of stateful and stateless protocol. For example, we can have a stateful protocol such as TCP running over stateless protocols such as IP, running over a stateful protocol such as LAPB. This depends entirely on the individual use case.

## B. Benign complications at the point of capture

Here we consider features of packet data that are to be anticipated, due to the way data is collected:

- **Defining a consistent flow tuple**: As described earlier, the concept of a flow depends on what features are included in the flow tuple, for which there are several standards, proprietary methods, and interpretations. To be useful we should align flow collection with what is achievable today in real collection systems, and could be extensible (e.g., IPFIX), particularly if we want to produce flow data as training data that could later be tested against production data.

- **Flow Direction and dynamics**: Connections typically fall into two types: half-duplex (one-way) and full-duplex (two-way). Connections can also be *inbound* or *outbound* with regard to the local network (which may not be obvious without details of the network address space). This requires additional logic to track flow state and deal with reversal of key identifiers in packets (addresses, ports etc.). This is important both for analytics and to maintain correct context for security analysis[8].

- **Partial Flow Capture**: This is a subtle problem that can have unanticipated side-effects. In a live network there is no natural 'start point' with which to start packet capture (unless we can orchestrate all networked devices to start simultaneously, which is highly unlikely outside a test environment). This means invariably that there will be many hundreds or even thousands of packet flows in progress at the time capture starts - see Fig.4. The same applies at the point of stopping the capture, where we may be missing flow termination exchanges[9]. The impact of this on flow recovery is described in *'Implications'*.

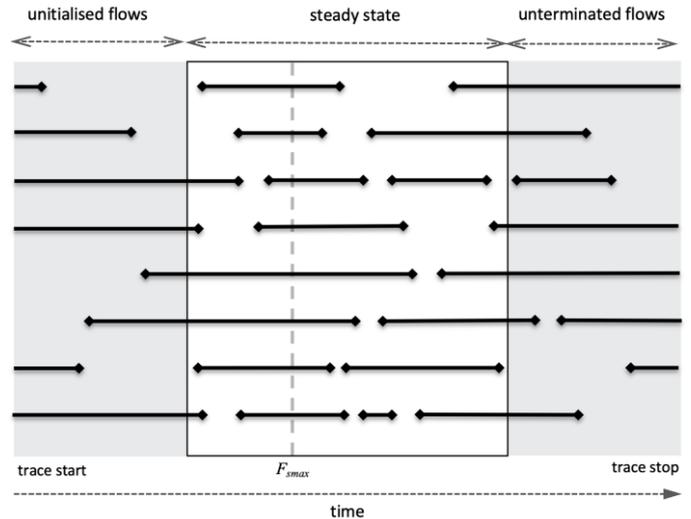

Fig. 4. Illustration of the three phases of a packet trace from a flow perspective. Initially there may be mainly uninitialised packet flows, where flow start sequence exchanges are missing, with the reverse is true at the end of the trace. To be clear, these flows are not actually uninitialised, but from the perspective of flow recovery they appear to be. In the middle phase the majority of flows will be complete exchanges, with session transition features intact and observable by the flow recovery method. The boundaries and presence of a *steady state* are somewhat arbitrary, and will depend on the particular use case. For example a short packet trace from a high bandwith network may contain only partial flows and no steady state. The concept of a steady state also depends on the notion of an acceptable ratio of flow states, since we are guaranteed that there will not be a particularly long uninitialised flow that extends throughout the entire trace. These phases are an artefact of how packet traces are taken, with defined start and stop points. In a live network we are likley to be processing flows continuously in real time, with a consistent steady state. Note that $F_{smax}$ is the 'deepest' point in the trace, where the maximum number of concurrent flows is in play concurrently.

## C. Adversarial and anomalous activity

Here we consider features of packet data that may or may not be anticipated, due to the way packet streams can be manipulated by adversaries. For this reason, the flow recovery technique should not make assumptions that everything in the packet trace will comply with standard or expected behaviour. These artefacts can significantly impact flow recovery design:

- **Protocol abuse**: bad actors do not care much for standards, other than to find ways to

---

[8] Direction matters in cybersecurity - knowing whether a flow originated from an external system or internal system can be critical information in deciding the motives and objectives of an attack.

[9] In our analysis of several popular datasets [KEN20] we observed significant levels of uninitialised TCP flows in the range from 2 to 16%, and unterminated flows in the range of 5 to 99%. In practice these ranges will vary widely depending on how the trace was captured and the deployment context.

manipulate them. In malicious traffic it should be expected that features such as flags, timers, buffers may be used in unexpected ways.

- **Protocol state machine abuse**: bad actors will attempt to manipulate resources by finding ways to leverage system and protocol behaviour. For example, a SYN Flood [BOGD13, SCHUBA17] is a volumetric attack that compromises the default TCP handshake conventions widely adopted across networked systems, where the final phase of the handshake never takes place, and the server is left holding valuable session memory open[10]. This is important in flow recovery because the partial state transition, and lack of a subsequent flow termination, need to be dealt handled robustly.

- **Non-terminating 'zombie flows'**: bad actors are increasingly finding ways to mount attacks that adhere to standard protocol behaviour, but in practice manipulate resources. For example, by starving systems of resources over extended periods with slow applications layer attacks (such as *Slowloris*) [SURO17]. In flow recovery these features may for example appear as normal (i.e., protocol compliant) flows that extend for very long periods, may never terminate, and must be held in memory for the duration. As shown in Fig.4, these are a special class of unterminated flow.

*D. Implications*

The challenges identified pose several problems for flow extraction, which leads to some counter-intuitive design problems. We cannot always rely on 'correct' flow and state machine behaviour due to:

- protocol implementation errors
- faulty equipment and misconfiguration
- missing or lost packets
- partial flow capture
- ambiguous flow termination
- malware attacks that manipulate weaknesses in state handling

For example, assuming all TCP flows start with a SYN flag exchange seems reasonable. However, an implementation of flow state recovery based on the need to observe a SYN exchange may fail, for several reasons:

1. Uninitialised flows may not contain this exchange, in which case the recovery process may inadvertently drop these flows without a more flexible heuristic. So, this is a problem for the flow recovery logic - for example in how flow state is used to determine memory caching and cache object assignment (see point 2 below). It is also a potential concern (not widely discussed in the literature) if flow data is to be used as training data. For example, in intrusion detection research - depending on the features used to predict anomalies - some traces may exhibit distortions in the distribution of features (e.g., if we were basing a classifier partly on the presence of TCP SYN flags). As noted earlier in Fig.4, flow collection is likely to be handled in real time, with a consistent steady state. We should therefore tailor training data according to the production context so that models training on such data perform well in production.

2. A more subtle problem with partially observed flows is that flow direction can be ambiguous, since based only on the first packet we do not know who initiated the flow. In such cases it is unclear whether the initial flow was *inbound* or *outbound*. Statistically we should expect to see half of such flows to be incorrectly assigned. This has implications for security analysis (noted earlier), and any ambiguity may also impact learning models if flow features are assigned incorrectly. For example, *destination* TCP/UDP ports generally have much more significance than the source port[11]. If ports are recorded incorrectly during flow recovery, then attempts to learn from the resulting flow data may be flawed - e.g., where we might correlate destination port with other features in a classifier.

---

[10] This is eventually terminated by a timer but the default times tend to be 30 or 60 seconds, which is a long time in networking, and with finite buffers and repeated attacks memory can be exhausted rapidly, at which point legitimate users are no longer able to connect.

[11] See [KEN02] for details of port assignments. The key point to note is that ports are assigned in three blocks: 1. well-known, 2. registered, 3. dynamic and/or private ports. The first two blocks are primarily used as standard destination identifiers when connecting *to* a service, using the destination port.

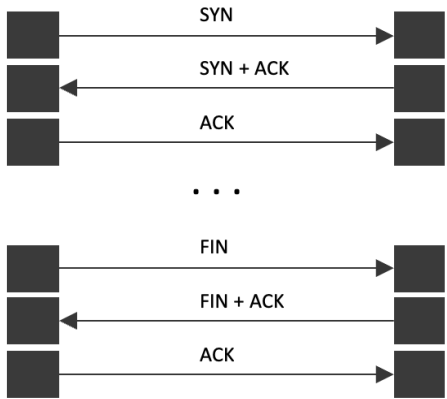

Fig. 5. TCP protocol flag states during a normal 3-way handshake to initiate a session (analogous to a flow) from source to destination, and during session termination. Note that during termination can be initiated by either party, and in real implementations it is possible that once the FIN flag is sent that the initiator assumes the connection is closed

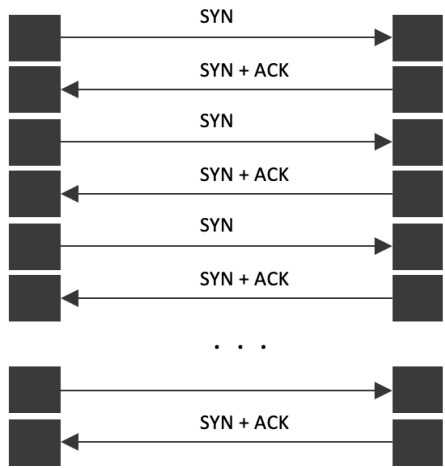

Fig. 6. TCP protocol flag states during a volumetric DDoS attack (a SYN Flood [SCHUBA97]). In order to capture such artefacts we assume that a new (i.e. currently unknown) flow tuple signifies a new flow is starting to avoid generating many thousands of flows wit just two packets. For dimensionality reduction and improved analysis in this case it is typically more convenient to aggregate all these packet exchanges into a single flow.

3. Well-behaved TCP flows observe consistent state transitions at the start and end of a flow - see Fig.5. Malicious flows, such as a SYN Flood [SCHUBA97], create ambiguities in how a flow should be initialised and tracked, and even what constitutes a discrete flow. For example, whether to create a separate flow instance for each SYN - SYN-ACK pair or aggregate all packets into a single flow - see Fig.6. This partly depends on whether the flow tuple is identical for a set of packets[12], packet

---

[12] Source network addresses are often spoofed in volumetric DDoS attacks, and attacks on a specific host may use many different source network addresses. However, based on packet timing proximity, the anomalous

timing proximity, and whether we can logically group these events based on common destination identifiers.

Based on the challenges described, flow recovery logic needs to be designed with the flexibility and robustness to deal with a wide array of unexpected behaviour, including missing data. It is also important to note that flow recovery should be broadly in line with what is *practically deployable in real-world flow capture*. For example, training data produced by the recovery process might have limited utility in real world applications if similar flow feature representations cannot be captured in live deployments (for example using IPFIX [RFC7011] based flow collectors).

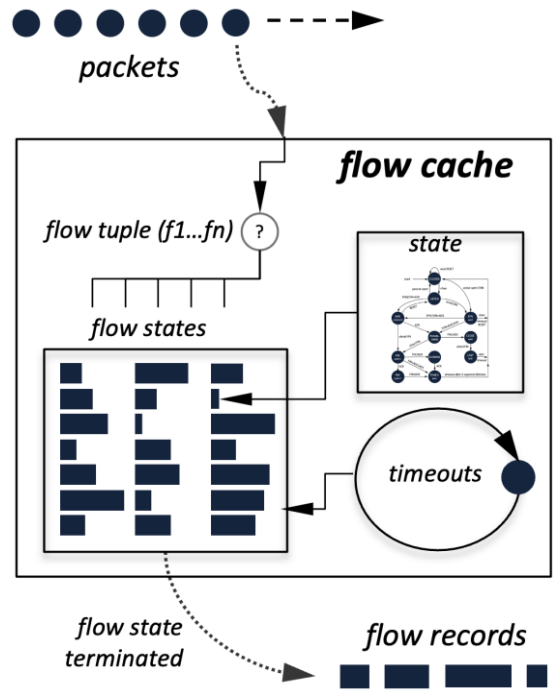

Fig. 7. The flow cache maintain state for all flow until termination conditions are met (through the appropriate protocol state transition, or a timeout). In practice there may be tens of thousands of flows in the cache at any time during processing of a large packet trace.

IV. POPOSED METHODS

We have implemented a proof-of-concept code to demonstrate how to mitigate these challenges [GSX]. Specifically, we focus on dealing with:

- Steady state

---

session exchange, and a common destination address and port, we could aggregate these packets into a common flow.

- Missing flow initialisation
- Flow direction ambiguity
- Flow sequencing issues

The main concepts are described below.

### A. Flow state memory cache

At the heart of our implementation is a **cache** to hold all flows in process. This is based on a fast lookup table to optimise performance (i.e., a hash table), using a flow tuple as the key. Fig.7 shows the key concepts in how flows are recorded as a packet trace (or incoming packet feed) is processed.

The main features of the flow and packet handling logic proceed as follows:

1. As a PCAP trace is read sequentially, we construct a flow tuple dynamically from each packet, and perform a cache lookup. Since flows are bi-directional, this lookup must be attempted in both directions since we hold one cache entry per flow.

2. If there is no cache hit, then a new flow data structure is created from the packet features and is added to the cache (flow direction is assigned by default but may be overridden if we can infer this - described shortly).

3. If there is a cache hit, then the flow direction is assigned, and the existing flow data structure referenced by the cache is updated (counters, statistics, timing, special features, state etc.).

4. For protocols such as TCP, flow state is dynamically tracked by observing flag state, and where session termination flags are observed, or flow timeouts expire, flows are removed from the cache and added to a *flow file* (in CSV format). This file is dynamically built during the recovery process. Each row represents an entire flow record, summarising all packet activity on that flow.

5. Flow termination can be triggered by a normal TCP termination sequence (using the FIN and ACKL flags), or inactivity timeouts.

6. Flow termination can also be triggered prematurely by receipt of a new SYN flag.

In this way we can observe cache entries building up to the point where the cache typically achieves a steady-state, as older flows age out and new flows are added. A summary of the flow handling logic is given in Fig.8.

It is important to note that the flow recovery process assumes that not all protocol state rules will be adhered to (for example, some attack types may violate both state flags and timing), and so the above logic (especially termination) accounts for such issues. There are also additional diagnostic counters to identify potential anomalies discovered in the recovery process. This implementation provides more reliable flow recovery, more accurate and useful flag representation, and more optimal input for learning models.

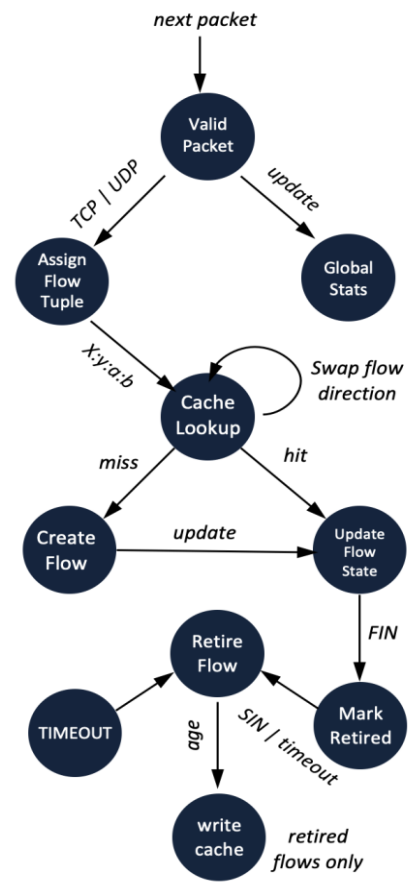

Fig. 8. Core logic in flow reconstruction. Note that the emphasis is (counter-intuitively) on the FIN flag. When we received a packet the flow tuple is used to decide whether the flow should be assigned or created - regardless if a SYN is present or not. This supports aggregation of volumetric attacks into a common flow (if the addresses are consistent) and also ensures that flows are properly terminated.

*B. Flow output sequencing*

As just described, flows that are dynamically terminated or aged from the flow cache, are moved into a flow file. Initially this file will be sorted by the order in which the flows are moved out of the cache, and so the flow start times will very likely be out of order (long lasting flows will stay in the cache much longer). The flow file can be resorted by the first packet arrival time once the flow file is complete.

*C. Cache memory size*

In large traces spanning many hours it is not unusual to observe tens of thousands of active flow states in memory, and therefore any implementation should have sufficient resources for this (the exact memory required will depend largely on the number and complexity of features being maintained for each flow, bearing in mind that flows are often bi-directional, so this be doubled, hence a worst-case estimate for cache size would be:

$F_{smax}$ x $(D_{so} + 2$ x $Fs)$

Where $F_{smax}$ is the maximum concurrent flow count (as illustrated in Fig.4[13]), $D_{so}$ is the data structure overhead within the flow data structure (flow tuple, any summary statistics etc.) and $Fs$ is the memory required for the feature subsets engineered during flow extraction (e.g., packet and byte counters, payload statistics). As a general rule of thumb, the ratio of packets to flows tends to be roughly 30:1[14] based on prior research [KEN20] and observations made during implementation. We could consider a worst-case scenario being a short timespan packet trace containing a volumetric DDoS attack where every two packets (corresponding to a TCP SYN and SYN-ACK) are assigned a flow entry, and all flows have not yet expired.

*D. Uninitialised flow handling*

As illustrated in Fig.4, the presence of flow initialisation is not necessary for flow recovery. The flow cache key is assigned based on a unique flow tuple[15]. Flow direction handling is discussed in the following section. Depending on the intended application, we may wish to use the whole packet trace for building a flow dataset, or a subset of 'cleaner' data (e.g., if we want to build a training dataset mostly composed of complete flow exchanges).

Analysis of the flow composition can reveal the location where a *steady state* is reached, as illustrated in Fig.4, though this will depend on the particular use case, and the ratio of full to partial flows that are to be considered to define the boundary (asymptotically this could be the point in time when only fully initialised flows are present in the trace, although we are not guaranteed to find such a state, particularly with shorter packet traces).

Given information on the position of a steady state, we could choose to extract PCAP samples from the original trace, starting at packet ID closest to the start of steady state, and then run flow recovery on that data. We might also decide to drop all flows that are not initialised (or terminated), which would effectively produce a clean flow dataset. These problems can be solved by making such actions configurable, so that the user can decide on the appropriate handling based on the specific application.

*E. Flow direction by inference*

In a number of public flow datasets [KEN20] we observed that flow direction was recorded incorrectly. We see for example that *well-known* or *registered* destination TCP and UDP ports [KEN02] are assigned as a *source port* feature. As we discussed earlier, the proportion of flows incorrectly assigned will vary by the dataset and deployment context[16].

To allocate flow direction correctly we should therefore try to *infer* the direction of the flow. This could be achieved in several ways, for example:

---

[13] As a worst case this might be equal to all flows, depending on the use case.
[14] Based on detailed analysis of several open datasets [KEN20], this may vary depending on the deployment context.
[15] Strictly speaking the tuple is not guaranteed to be unique over time with TCP/IP, since ephemeral ports numbers will eventually wrap around once they reach the maximum range. However, unless network equipment is faulty, this process typically takes a long time and flows can be easily separated either by parsing the flow termination events or using reasonable timeouts.
[16] For example, short duration, large packet traces, on high bandwidth links may have a higher proportion of uninitialised flows than a day long packet trace from an enterprise network, where a steady state may have been reached for several hours.

- **Known Address Maps**: Using additional contextual address data for the network infrastructure - for example, identifying all internal address ranges and subnets. This would require this data to be made available to the flow recovery process and used as a lookup table during flow recovery.
- **Inferred Port Assignment**: Dynamically inferring direction based on ports assigned in each packet, against known port number ranges.

Since we cannot assume that infrastructure information is available, and we can reasonably expect most service port numbers would be associated with the destination port[17], we adopt the latter approach[18]. The *Flow Direction Inference* feature has been implemented so that even where a flow is only partially available, on receipt of the first packet, an attempt is made to infer flow direction based on the presence of a recognised service port identifier.

To optimise this process further, we make the assumption that the *lowest* numeric port is generally going to be the most important and should be expected to be either in the *well-known* or *registered* IANA (Internet Assigned Numbers Authority) ranges[19]. We can then infer flow direction and assign the most likely destination port (flipping all other packet features accordingly where needed)

Where flow direction cannot be inferred the flow will be assumed to be in the forward direction, which should be expected to balance out statistically over a large sample size (roughly half of these flows should be correctly assigned). In experiments it was found that intelligent flow recovery accounted for as much as 20% of flows in some packet traces - and we can reasonably assume half of these would otherwise have been misclassified by other tools.

*F. Special case handling*

Our implementation also accounts for some notable exceptions (such as FTP 'well-known' control and data ports, which are in opposing directions). Note that there are some protocols (such as IPSec [RFC4301, RFC4309]) where port allocation may be much more complex and dynamic, and this may require deeper packet inspection of packet payload (assuming it is not encrypted) to unpack these port changes and capture the flow in its entirety.

*G. Flow termination handling*

In several public datasets we observed anomalies in flow state counters for TCP FIN flags - specifically these counters were either zero or did not balance the number of SYN flags observed[20]. This points to issues in either correctly terminating flow states in the flow cache or terminating the flow too early (which would potentially result in seeing 'late' packets after the cache entry has been flushed, and possibly creating new partial or broken flows, or discarding those packets (which means that flow packet counters are not strictly correct either). Additional state handling has been implemented to prevent ambiguity in flow retirement. Essentially there is a 'wait period' where a flow is held in the cache until all final interaction has taken observed, even after a FIN has been seen or an inactivity timer expires. Flows are not retired as soon as a FIN is observed.

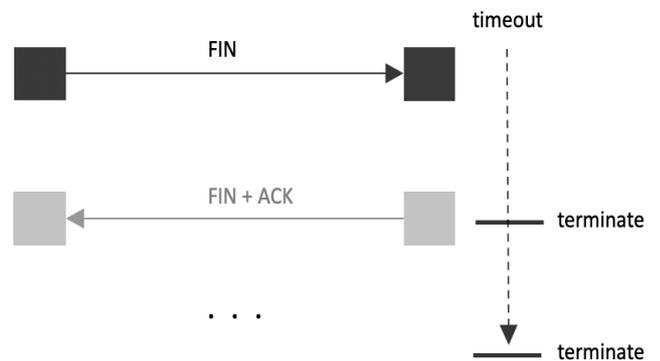

Fig. 9. Ambiguity with TCP reconstruction in flow termination. TCP flow termination, where we wait for either a corresponding FIN back from the reciever or rely on a timeout. Whilst not perfect this metod does associate FIN flags more accurately with flows, provides accurate FIN counts, and

---

[17] We must exclude some ports such as FTP and FTPS data ports since the direction in these cases is reversed.

[18] Additionally, we could attempt to infer information on local and external addresses based on deep packet inspection of protocols such as ARP, DNS, DHCP, and routing protocols.

[19] It is possible that a proprietary implementation or configuration might break this assumption where both ports are technically in the dynamically assigned range, however in that case we will still need additional contextual information to be able to differentiate this scenario.

[20] It is reasonable to assume that in a large enough set of packet samples we should see balanced counts for TCP SYN and FIN flag, assuming that we have a fully observed flows.

does not overcount flows. Depending on the implementation some TCP end points may behave differently at the point of closing down a TCP session

### H. Implementation details

Our proof of concept is implemented in Python, using the Scapy package for packet inspection support [SCAPY23] - or further details on implementation see [GSX].

**Dealing with large file sizes**: Given that the type of network packet traces used in network and cybersecurity analysis can often be hundreds of gigabytes in size, special attention is required in memory management. In naive implementations packet processing logic is often implemented so that the entire packet trace is loaded into memory, which has obvious processing speed advantages. In practice these packet traces tend to be too large to fit in memory, and so we implement packet processing by reading PCAP trace files iteratively, at the event level. Cache entries are implemented as *flow objects* that can be updated dynamically during processing.

It is also worth noting that some public trace datasets span multiple days. It can be valuable to merge these traces during the flow recovery process, which is relatively trivial to implement since we read events iteratively. The value of this feature depends on whether packet traces are contiguous (for example if file rotation is implemented during packet capture, and there are no packets lost between traces). If so, then it is possible to continue flow processing across packet traces by simply switching the current file handle. In this case flow entries will persist in the flow cache and remain valid for consecutive packet traces.[21]

**Diagnostics**: As discussed earlier, flow recovery is not an exact science, and in our analysis [KEN20] we found that public flow traces occasionally have minor formatting issues - which required additional pre-processing (for example, hexadecimal fields appearing in numeric columns, delimiter formatting issues). In addition to this some flows exhibit anomalous characteristics at recovery time (for example, TCP SYN flags appearing after the flow has terminated). To support the recovery process, additional diagnostic metadata is therefore included with each flow. This can be used to assess flow fidelity with respect to protocol state handling, distributions, and others useful characteristics (for example by recording late packet arrivals after a flow has been terminated). This helps support objective assessment of the quality of synthetic output qualitatively and help identify anomalies in flow states or the raw packet data.

**Label Recovery**: Datasets may include embedded labels as features within the flow record set, for example those used for intrusion detection [KEN20]. Where flow records are to be reconstructed from the original packet data, it may be important to recover and re-apply these labels, particularly where the new dataset is to be used in machine learning. In practice label recovery can be problematic. If there are flaws in the original flow recovery method, the resulting flow datasets may differ in several ways.

- The number of flows may not match, for example where the criteria or accuracy in aggregating packets into flows differs.
- Flow record order may differ, due to differences in the flow state tracking and cache behaviour (in which case flow position and flow order may differ).
- Flow duration may differ with some flow types. For example, where the flow recovery method does not track TCP flow state termination fully, the flow duration may be prematurely short.
- In some cases, we observed that flow timestamps and packet timestamps may be inconsistent, may be offset by several hours, or may be inconsistent between adjacent record sets (in a multi-trace archive).

Through experimentation we found the most reliable way to re-apply labels to a new flow dataset is to follow the procedure below:

1. **Apply default labels**: If there is a default label annotation (e.g., 'normal'), apply that to all samples in the new flow dataset as the default. sample state

---

[21] If there are time gaps and packets missing between successive packet traces, then the value is less clear. Although it is possible that some long running flows may still span multiple traces, there will be missing data in those flows, and potentially misleading flow state and statistics.

2. **Prepare a signature dataset**: Extract a complete set of signatures[22], per flow sample. We used a signature tuple, comprising timestamp, flow duration, protocol, network address pair, port number pair, and label assignment.
3. **Re-apply signatures**: Iterate through the new flow dataset, a) performing a search for signature matches on *protocol*, *address* and *port pairs*, then b) Perform a weighted match against the search results, using timestamp and flow duration. This second step differentiates matches where the search finds multiple records. c) Apply the label only where there is a high corelation between the signature and the flow record. d) Maintain a list of errors for unmatched signatures, together with the flow Id (or sample row number).

In practice it was necessary to relax the tolerance of exact timestamp and flow duration matches, for reasons described earlier. The method described here works well in practice, with negligeable errors with good explainability, although it could be extended to include additional features in the tuple match - the main issues found tend to be mismatches in flow duration and timestamps with specific flow record sets.

V. DATASET COMPOSITION

As part of this work, we examined novel ways to view the composition of packet datasets, using data extracted during flow recovery. This included:

- Visualisation of peer connectivity graphs
- Visualisation of flow state summaries
- The extraction of flow summary statistics
- Advanced flow feature engineering, including novel features such as payload entropy.

Detailed discussion of each of these is outside the scope of this paper, however we describe one method below that can provide useful insights on dataset composition, from the perspective of *flow state*.

---
[22] For those flow record samples associated with non-default labels (e.g., samples that are not labelled as 'normal', such as anomalies, malware indicators etc.).

A. *State Symbol Sequence Generation*

As part of the proof of concept we included the ability to extract TCP state flag sequences and export these in compressed symbolic form to text files. Fig.10 shows a high-level example where we track TCP protocol state flags by flow and by direction (note that this could be easily extended, or even converted into a graphical form). For example, in the case of a TCP flow we would expect a sequence of state flags similar to the one below[23] (note that the symbols in uppercase indicate 'forward' direction, lowercase indicate reverse):

S S sa sa A A PA PA a pa pa A A PA PA a a pa PA ra ra r r FA FA

This sequence represents a complete TCP flow lifecycle. Flags are exported in compressed form, where S=SYN, A=ACK, F=FIN, P=PSH, U=URG, for all flows recovered from the packet trace, where we might expect to see something resembling the flow sequence data below (flows are delimited by a period '.' symbol):

```
. S S sa sa A A PA PA a pa pa A A PA PA a a pa pa PA PA pa pa PA PA pa pa A A PA PA ra ra
. S S sa sa A A PA PA a pa pa A A FA FA a a ra ra . S S sa sa A A PA PA a a pa pa PA PA
pa pa PA PA pa pa PA PA ra FA ra FA r r . S sa S sa A A PA PA pa pa FA FA a a ra ra . S S
sa sa a a A A A PA PA pa pa FA FA a a ra ra . S S sa sa A A PA PA pa pa A A A a a pa PA
pa PA pa pa PA PA pa pa A A PA PA pa pa PA PA pa pa A A PA PA pa pa PA PA pa pa PA PA
pa pa PA PA pa pa A A PA PA pa pa PA PA pa pa PA PA pa pa A A RA RA . S S
sa sa A A PA a PA a pa pa A a A FA FA a ra ra . S S sa sa A A PA pa FA FA a a ra ra
. S sa A PA a pa fa A FA a . S S sa sa A A PA PA a a pa pa A A FA FA a a ra ra . S S sa sa
A A PA PA a a pa pa PA PA pa pa PA PA pa pa A A PA PA FA ra ra FA r r . S S sa sa A A PA
PA a a pa pa PA PA pa pa PA PA pa pa A A PA FA FA ra r r . S S sa sa A A PA PA pa pa A A
PA pa pa A PA PA a a pa pa PA PA pa pa PA PA pa pa PA PA pa pa PA PA pa pa A A
PA PA PA pa pa pa A A RA RA . S S sa sa A A PA PA pa pa PA PA pa pa PA PA pa pa PA
PA pa pa A A PA PA PA PA pa pa pa A A RA RA . S S sa sa A A PA PA pa pa PA PA pa pa FA fa
fa A A . S sa A PA a pa PA pa pa A . S S sa sa A A PA PA pa pa A A PA FA FA a a ra ra
. S S sa sa A A PA PA a pa pa A A PA PA pa pa A A PA PA pa pa PA PA FA FA ra
ra . S S sa sa A A PA PA a a pa pa A A FA FA a ra a ra . S sa S sa A A PA PA pa pa A A PA pa
PA pa pa PA PA pa pa PA PA pa pa PA PA pa pa A A PA PA pa pa A A PA PA pa
pa A A PA PA pa pa PA PA pa pa A A PA PA pa pa A A PA PA pa pa ra ra A r A
r . S S sa sa A A PA PA a a pa pa A A FA FA a a ra ra . S S sa sa A A PA PA pa pa A A PA
```

Fig. 10. Example of flow state symbol export with TCP state flags highlighted. (S=SYN, A=ACK, F=FIN, P=PSH, U=URG). Lower case symbols indicates reverse packet flow.

This provides us with a novel way to visualise the flow composition with respect to time, as a *stratified* view, with the oldest flows on top. One of the most striking aspects is the ability to see where a 'steady state' is achieved, and where new flows are being established or torn down in large clusters.

- What we typically see at the start of a large packet trace is a high proportion of flows where acknowledgements are being sent back and forth, and where no flow session initiation flags are observed. This can be useful in deciding where to start the flow recovery process when constructing training sets (for example if we do not want to deal with the noise associated with partial flows).

---
[23] In practice TCP flows tend to be longer, often with periods of inactivity, being kept 'alive' with bidirectional acknowledgements.

- This type of symbolic output may also be useful for use with machine learning techniques that can memorise symbolic sequences. As an example, we might use an RNN for example to learn and predict state sequences in various protocols (this could be useful for example in assisting the generation of synthetic time-series data).

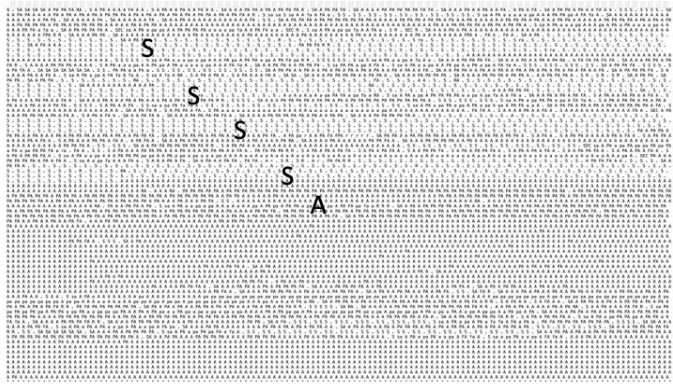

(a)

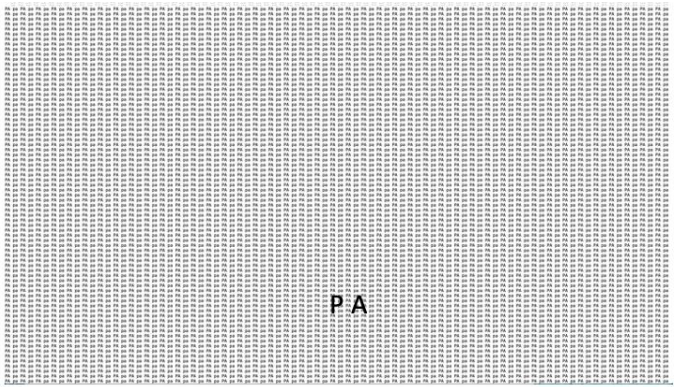

(b)

Fig. 11. Example of flow state snapshots from different layers of a flow dataset, where (a) appears near the start of the trace, (b) appears later in a trace). Flag groupings are highlighted by S=SYN, A=ACK, P=PSH. In (a) we see what appear to be a number of connection requests starting in clusters. In (b) we see a long phase of very stable open connections.

Fig.11 shows example flow symbol summaries from different trace positions where we can clearly make out coordinated bursts of connection requests (denoted by the SYN flag, in Fig.11.a), and a corresponding steady state in Fig.11.b, where only keepalives and data are being transmitted in established flows.

### B. Dimensionality reduction - big data verses small data feature extraction

Our implementation offers the ability to export a complete feature set (big data approach) or specific subsets (small data approach), including the ability to set thresholds on filtering highly correlated features. The 'small data' approach is likely to produce faster convergence in learning and detection models, but the decision on the scope of the feature set used should be determined by the researcher, and depends on factors such as:

- Time demands placed on detection in a deployment context (e.g., do we need near real-time detection).
- Resource constraints (memory, compute, storage).
- Whether a *broad* threat and anomaly detection is being researched, or *discrete* threat types are to be analysed (the most predictive feature sets are likely to differ).

It is worth noting that there could be variations in the importance of some features, depending on the deployment context, and whether address information has been anonymised or obfuscated in the dataset. For example, in the literature we often see *IP address* features omitted, since these are elements of the flow tuple and therefore implicit in the flow. In a real-world deployment certain IP addresses could be highly significant in a threat detection - for example servers that contain Personally Identifiable Information (PII). Note that a complete description of feature engineering is outside the scope of this paper, since we focus here on the problems in *flow record* recovery, further details are provided at [GSX].

### VI. CONCLUSIONS

Packet flow datasets are an important tool in the fight against cyber threats and can provide invaluable training data to help improve modern threat detection techniques. Flow datasets provide a lower dimensional view of packet data, and with careful engineered feature sets network traffic can be processed efficiently for deep behavioural analysis and predictive analytics. Generating accurate packet flow data post-facto from raw packet traces might at first seem relatively straightforward. However, as we discussed, there are a number of issues that make this task challenging, and mistakes in flow recovery can compromise analysis and distort important feature correlations For example, where important features are under-counted or wrongly assigned, this has the potential to distort feature importance ranking when performing anomaly detection. In this paper we outlined the core challenges, and proposed methods to mitigate these issues, including flexible flow state logic, port inference, partial flow tracking, and more relaxed flow termination logic.

To support this work, proof of concept code[24] has been made available as open source on GitHub [GSX]; please subscribe the repository for updates. We hope that some of these additional features will prove useful to researchers in improving detection techniques and dataset generation.

## VII. FURTHER WORK

One of the areas explored in this paper was the possibility of dataset analysis and visualisation. Given the size of these datasets, the complexity of the underlying data and class. distributions, gaining more insights on the composition and content of these datasets, particularly using visual techniques, is an interesting prospect to explore further. Further, we briefly touched on the potential for deep feature engineering, and this is a topic for additional publications to follow, such as the work on payload entropy cited in [KEN23E].

---

[24] This code focusses primarily on the TCP/IP suite, but can be extended.